\documentclass[runningheads]{svmult}

\usepackage{graphicx}  
\usepackage{subeqnar}  
\usepackage{cropmark} 
\usepackage{physprbb}  

\newcommand{\beq}{\begin{equation}}
\newcommand{\eeq}{\end{equation}}
\newcommand{\bea}{\begin{eqnarray}}
\newcommand{\eea}{\end{eqnarray}}
\newbox\hdbox%
\newcount\hdrows%
\newcount\multispancount%
\newcount\ncase%
\newcount\ncols
\newcount\nrows%
\newcount\nspan%
\newcount\ntemp%
\newdimen\hdsize%
\newdimen\newhdsize%
\newdimen\parasize%
\newdimen\spreadwidth%
\newdimen\thicksize%
\newdimen\thinsize%
\newdimen\tablewidth%
\newif\ifcentertables%
\newif\ifendsize%
\newif\iffirstrow%
\newif\iftableinfo%
\newtoks\dbt%
\newtoks\hdtks%
\newtoks\savetks%
\newtoks\tableLETtokens%
\newtoks\tabletokens%
\newtoks\widthspec%
\immediate\write15{%
CP SMSG GJMSINK TEXTABLE --> TABLE MACROS V. 851121 JOB = \jobname%
}%
\tableinfotrue%
\catcode`\@=11
\def\tstrut{\vrule height3.1ex depth1.2ex width0pt}%
\def\and{\char`\&}
\def\tablerule{\noalign{\hrule height\thinsize depth0pt}}%
\def\thickrule{\noalign{\hrule height\thicksize depth0pt}}%
\def\ctr#1{\hfil\ #1\hfil}%
\tablewidth=-\maxdimen%
\spreadwidth=-\maxdimen%
\def\tabskipglue{0pt plus 1fil minus 1fil}%
\centertablestrue%
\gdef\ARGS{########}
\gdef\headerARGS{####}
\def\@mpersand{&}
{\catcode`\|=13
\gdef\letbarzero{\let|0}
\gdef\letbartab{\def|{&&}}%
\gdef\letvbbar{\let\vb|}%
}
{\catcode`\&=4
\def\ampskip{&\omit\hfil&}
\catcode`\&=13
\let&0
\xdef\letampskip{\def&{\ampskip}}%
\gdef\letnovbamp{\let\novb&\let\tab&}
}
\def\begintable{
   \begingroup%
   \catcode`\|=13\letbartab\letvbbar%
   \catcode`\&=13\letampskip\letnovbamp%
   \def\multispan##1{
      \omit \mscount##1%
      \multiply\mscount\tw@\advance\mscount\m@ne%
      \loop\ifnum\mscount>\@ne \sp@n\repeat%
   }
   \def\|{%
      &\omit\widevline&%
   }%
   \ruledtable
}
\long\def\ruledtable#1\endtable{%
   \offinterlineskip
   \tabskip 0pt
   \def\widevline{\vrule width\thicksize}
   \def\endrow{\@mpersand\omit\hfil\crnorm\@mpersand}%
   \def\crthick{\@mpersand\crnorm\thickrule\@mpersand}%
   \def\crthickneg##1{\@mpersand\crnorm\thickrule
          \noalign{{\skip0=##1\vskip-\skip0}}\@mpersand}%
   \def\crnorule{\@mpersand\crnorm\@mpersand}%
   \def\crnoruleneg##1{\@mpersand\crnorm
          \noalign{{\skip0=##1\vskip-\skip0}}\@mpersand}%
   \let\nr=\crnorule
   \def\endtable{\@mpersand\crnorm\thickrule}%
   \let\crnorm=\cr
   \edef\cr{\@mpersand\crnorm\tablerule\@mpersand}%
   \def\crneg##1{\@mpersand\crnorm\tablerule
          \noalign{{\skip0=##1\vskip-\skip0}}\@mpersand}%
   \let\ctneg=\crthickneg
   \let\nrneg=\crnoruleneg
   \the\tableLETtokens
   \tabletokens={&#1}
   \countROWS\tabletokens\into\nrows%
   \countCOLS\tabletokens\into\ncols%
   \advance\ncols by -1%
   \divide\ncols by 2%
   \advance\nrows by 1%
   \iftableinfo %
      \immediate\write16{[Nrows=\the\nrows, Ncols=\the\ncols]}%
   \fi%
   \ifcentertables
      \ifhmode \par\fi
      \hbox to \hsize{
      \hss
   \else %
      \hbox{%
   \fi
      \vbox{%
         \makePREAMBLE{\the\ncols}
         \edef\next{\preamble}
         \let\preamble=\next
         \makeTABLE{\preamble}{\tabletokens}
      }
      \ifcentertables \hss}\else }\fi
   \endgroup
   \tablewidth=-\maxdimen
   \spreadwidth=-\maxdimen
}
\def\makeTABLE#1#2{
   {
   \let\ifmath0
   \let\header0
   \let\multispan0
   \ncase=0%
   \ifdim\tablewidth>-\maxdimen \ncase=1\fi%
   \ifdim\spreadwidth>-\maxdimen \ncase=2\fi%
   \relax
   \ifcase\ncase %
      \widthspec={}%
   \or %
      \widthspec=\expandafter{\expandafter t\expandafter o%
                 \the\tablewidth}%
   \else %
      \widthspec=\expandafter{\expandafter s\expandafter p\expandafter r%
                 \expandafter e\expandafter a\expandafter d%
                 \the\spreadwidth}%
   \fi %
   \xdef\next{
      \halign\the\widthspec{%
      #1
      \noalign{\hrule height\thicksize depth0pt}
      \the#2\endtable
      }
   }
   }
   \next
}
\def\makePREAMBLE#1{
   \ncols=#1
   \begingroup
   \let\ARGS=0
   \edef\xtp{\widevline\ARGS\tabskip\tabskipglue%
   &\ctr{\ARGS}\tstrut}
   \advance\ncols by -1
   \loop
      \ifnum\ncols>0 %
      \advance\ncols by -1%
      \edef\xtp{\xtp&\vrule width\thinsize\ARGS&\ctr{\ARGS}}%
   \repeat
   \xdef\preamble{\xtp&\widevline\ARGS\tabskip0pt%
   \crnorm}
   \endgroup
}
\def\countROWS#1\into#2{
   \let\countREGISTER=#2%
   \countREGISTER=0%
   \expandafter\ROWcount\the#1\endcount%
}%
\def\ROWcount{%
   \afterassignment\subROWcount\let\next= %
}%
\def\subROWcount{%
   \ifx\next\endcount %
      \let\next=\relax%
   \else%
      \ncase=0%
      \ifx\next\cr %
         \global\advance\countREGISTER by 1%
         \ncase=0%
      \fi%
      \ifx\next\endrow %
         \global\advance\countREGISTER by 1%
         \ncase=0%
      \fi%
      \ifx\next\crthick %
         \global\advance\countREGISTER by 1%
         \ncase=0%
      \fi%
      \ifx\next\crnorule %
         \global\advance\countREGISTER by 1%
         \ncase=0%
      \fi%
      \ifx\next\crthickneg %
         \global\advance\countREGISTER by 1%
         \ncase=0%
      \fi%
      \ifx\next\crnoruleneg %
         \global\advance\countREGISTER by 1%
         \ncase=0%
      \fi%
      \ifx\next\crneg %
         \global\advance\countREGISTER by 1%
         \ncase=0%
      \fi%
      \ifx\next\header %
         \ncase=1%
      \fi%
      \relax%
      \ifcase\ncase %
         \let\next\ROWcount%
      \or %
         \let\next\argROWskip%
      \else %
      \fi%
   \fi%
   \next%
}
\def\counthdROWS#1\into#2{%
\dvr{10}%
   \let\countREGISTER=#2%
   \countREGISTER=0%
\dvr{11}%
\dvr{13}%
   \expandafter\hdROWcount\the#1\endcount%
\dvr{12}%
}%
\def\hdROWcount{%
   \afterassignment\subhdROWcount\let\next= %
}%
\def\subhdROWcount{%
   \ifx\next\endcount %
      \let\next=\relax%
   \else%
      \ncase=0%
      \ifx\next\cr %
         \global\advance\countREGISTER by 1%
         \ncase=0%
      \fi%
      \ifx\next\endrow %
         \global\advance\countREGISTER by 1%
         \ncase=0%
      \fi%
      \ifx\next\crthick %
         \global\advance\countREGISTER by 1%
         \ncase=0%
      \fi%
      \ifx\next\crnorule %
         \global\advance\countREGISTER by 1%
         \ncase=0%
      \fi%
      \ifx\next\header %
         \ncase=1%
      \fi%
\relax%
      \ifcase\ncase %
         \let\next\hdROWcount%
      \or%
         \let\next\arghdROWskip%
      \else %
      \fi%
   \fi%
   \next%
}%
{\catcode`\|=13\letbartab
\gdef\countCOLS#1\into#2{%
   \let\countREGISTER=#2%
   \global\countREGISTER=0%
   \global\multispancount=0%
   \global\firstrowtrue
   \expandafter\COLcount\the#1\endcount%
   \global\advance\countREGISTER by 3%
   \global\advance\countREGISTER by -\multispancount
}%
\gdef\COLcount{%
   \afterassignment\subCOLcount\let\next= %
}%
{\catcode`\&=13%
\gdef\subCOLcount{%
   \ifx\next\endcount %
      \let\next=\relax%
   \else%
      \ncase=0%
      \iffirstrow
         \ifx\next& %
            \global\advance\countREGISTER by 2%
            \ncase=0%
         \fi%
         \ifx\next\span %
            \global\advance\countREGISTER by 1%
            \ncase=0%
         \fi%
         \ifx\next| %
            \global\advance\countREGISTER by 2%
            \ncase=0%
         \fi
         \ifx\next\|
            \global\advance\countREGISTER by 2%
            \ncase=0%
         \fi
         \ifx\next\multispan
            \ncase=1%
            \global\advance\multispancount by 1%
         \fi
         \ifx\next\header
            \ncase=2%
         \fi
         \ifx\next\cr       \global\firstrowfalse \fi
         \ifx\next\endrow   \global\firstrowfalse \fi
         \ifx\next\crthick  \global\firstrowfalse \fi
         \ifx\next\crnorule \global\firstrowfalse \fi
         \ifx\next\crnoruleneg \global\firstrowfalse \fi
         \ifx\next\crthickneg  \global\firstrowfalse \fi
         \ifx\next\crneg       \global\firstrowfalse \fi
      \fi
\relax
      \ifcase\ncase %
         \let\next\COLcount%
      \or %
         \let\next\spancount%
      \or %
         \let\next\argCOLskip%
      \else %
      \fi %
   \fi%
   \next%
}%
\gdef\argROWskip#1{%
   \let\next\ROWcount \next%
}
\gdef\arghdROWskip#1{%
   \let\next\ROWcount \next%
}
\gdef\argCOLskip#1{%
   \let\next\COLcount \next%
}
}
}
\def\spancount#1{
   \nspan=#1\multiply\nspan by 2\advance\nspan by -1%
   \global\advance \countREGISTER by \nspan
   \let\next\COLcount \next}%
\def\dvr#1{\relax}%
\def\header#1{%
\dvr{1}{\let\cr=\@mpersand%
\hdtks={#1}%
\counthdROWS\hdtks\into\hdrows%
\advance\hdrows by 1%
\ifnum\hdrows=0 \hdrows=1 \fi%
\dvr{5}\makehdPREAMBLE{\the\hdrows}%
\dvr{6}\getHDdimen{#1}%
{\parindent=0pt\hsize=\hdsize{\let\ifmath0%
\xdef\next{\valign{\headerpreamble #1\crnorm}}}\dvr{7}\next\dvr{8}%
}%
}\dvr{2}}
\def\makehdPREAMBLE#1{
\dvr{3}%
\hdrows=#1
{
\let\headerARGS=0%
\let\cr=\crnorm%
\edef\xtp{\vfil\hfil\hbox{\headerARGS}\hfil\vfil}%
\advance\hdrows by -1
\loop
\ifnum\hdrows>0%
\advance\hdrows by -1%
\edef\xtp{\xtp&\vfil\hfil\hbox{\headerARGS}\hfil\vfil}%
\repeat%
\xdef\headerpreamble{\xtp\crcr}%
}
\dvr{4}}
\def\getHDdimen#1{%
\hdsize=0pt%
\getsize#1\cr\end\cr%
}
\def\getsize#1\cr{%
\endsizefalse\savetks={#1}%
\expandafter\lookend\the\savetks\cr%
\relax \ifendsize \let\next\relax \else%
\setbox\hdbox=\hbox{#1}\newhdsize=1.0\wd\hdbox%
\ifdim\newhdsize>\hdsize \hdsize=\newhdsize \fi%
\let\next\getsize \fi%
\next%
}%
\def\lookend{\afterassignment\sublookend\let\looknext= }%
\def\sublookend{\relax%
\ifx\looknext\cr %
\let\looknext\relax \else %
   \relax
   \ifx\looknext\end \global\endsizetrue \fi%
   \let\looknext=\lookend%
    \fi \looknext%
}%
\def\tablelet#1{%
   \tableLETtokens=\expandafter{\the\tableLETtokens #1}%
}%
\catcode`\@=12
 
\begin{document}
\title*{ 
    Hadronic collisions: physics, models and event generators used for
    simulating the cosmic ray cascade at the highest energies
}
 \toctitle{Hadronic collisions: physics, models and event generators
}
\titlerunning{Hadronic collisions}
%

 \author{  J.~Ranft\inst{1} }
\authorrunning{J.~Ranft}
  \institute{
 Physics Dept. Universit\"at Siegen, D--57068 Siegen, Germany,
 e--mail: Johannes.Ranft@cern.ch}
\maketitle              
  \vspace{2mm}
 {\bf Siegen preprint SI--01--01}\\
 \vspace{2mm}
 \centerline{\it Extended version of a paper to be published in the}
\centerline{\it proceedings of the meeting Monte Carlo 2000, Lisboa, Oct. 2000}
 \vspace{2mm}

\section{Introduction}

Models and event generators for hadronic collisions belong to the basic
building blocks of hadron cascade calculations to simulate radiation
problems around high energy accelerators and to simulate the Cosmic Ray
cascade in the atmosphere, within the earth or within space stations. 
At the same time these
models often have to describe also 
the new physics to be investigated at a given
accelerator, examples are the present study of photon--photon collisions
at the LEP collider or the study of 
central collisions of heavy ions at the CERN--SPS and at the RHIC and
LHC colliders.

There are at least two main applications of such models to study
radiation problems around
accelerators and colliders:

(i)At colliders like the LHC one of the main radiation sources are the
proton--proton or nucleus--nucleus collisions inside the detectors.
Present and future colliders are also to investigate $e^+$ -- $e^-$, 
$\mu^+$ -- $\mu^-$  and photon--photon collisions. Therefore, 
also models for
hadronic collisions of leptons and photons become more and more
important for the study of radiation problems.

(ii)The second application is for the transport of the produced
particles in the elements of the detectors and accelerators and in the
shielding material. For this we have to understand hadron--nucleus,
nucleus--nucleus, lepton--nucleus, photon--nucleus and even
neutrino--nucleus collisions at all energies.

Here we are concerned with hadronic models or event generators for
hadronic collisions. These models are inserted into hadron cascade codes
to study the radiation transport. We will not discuss hadron cascade
codes here, but some of them are mentioned when discussing the
application of the hadronic models. There are also cases, where a
dedicated hadronic model is formulated within the hadron cascade code.

The models for hadronic collisions have to provide the produced hadrons
preferably in the form of Monte Carlo events as well as the inelastic 
cross sections of the collisions considered.
The event generators of interest for our applications have to provide
minimum bias events, not only events for selected (hard) collisions,
which might be the main object of the physics studied at a given
collider. We need detailed models for all collisions, not only for
interesting events, where for instance SUSY particles or Higgs bosons
might be produced.
In recent years new features of hadronic collisions have been found
experimentally which were not predicted by any of the models.
Therefore, the models
have to be continuously updated and revised in order to reproduce all
known properties of the collisions. Some of these features are:
(i)Hard diffraction \cite{Ingelman85,Bonino88} 
and events with large rapidity gaps between jets \cite{CDF98,Abbott98}.
(ii)Enhanced stopping of leading baryons in nuclear collisions
\cite{NA35FIN,Alber98},  one way to incorporate this into
  the multi--string models is by considering
new string configurations (diquark--breaking diagrams) 
\cite{Ranft20001,Ranft20003}.

\section{Models for hadronic interactions}

Unfortunately, it is not possible to discuss here all existing Monte
Carlo models for hadron production. Let me at least mention some
important models, which I will not discuss:

(i)There are JETSET \cite{JETSET}, PYTHIA \cite{Sjostrand93a}, HERWIG
\cite{HERWIG} and ISAJET \cite{ISAJET748}, which are mainly used to
sample selected classes of events. These are
certainly the  event generators used most extensively by collider
experiments. Codes not discussed, mainly for reasons of space, are also
HIJING \cite{Wang91} and FRITIOF \cite{FRITIOF}, both are codes for
nuclear collisions.

(ii)There are dedicated event generators inside hadronic  codes
like FLUKA, HERMES, MARS, MCNPX, NMTC  or GEANT. 
Usually these are not available as separate codes.
The event generators inside all of these codes will be covered by
contributions to this Meeting (examples: FLUKA: \cite{Sala001} GEANT:
\cite{Wellisch001}). Therefore, there is no need for me to discuss them.

(iii)PHOJET \cite{Engel95a,Engel95d} is a minimum bias Dual Parton Model
event generator with a very broad
applicability for hadrons, photons and electrons as projectiles and
targets. PHOJET is contained in the new DPMJET--III event generator,
which adds nuclei as targets and projectiles to the above PHOJET list.
DPMJET--III will for the first time be described at this meeting
\cite{Roesler20001}.  

In this talk I will only discuss  hadronic models 
which work above a 
mimimum lab--energy of about 5 GeV per nucleon and with a maximum 
lab--energy of about 10$^{11}$ GeV or (in the units used in Cosmic Ray 
applications 10$^{21}$ eV), 
this means codes which are also available for
studying the Cosmic Ray cascade up to the highest energy.

The extension of models for multiparticle production in hadron--hadron,
 hadron--nucleus and nucleus--nucleus collisions to be used for 
the simulation of the Cosmic Ray cascade 
up to $E_{lab}$ = 10${}^{21}$ eV
(corresponding  for p--p collisions to $\sqrt s $ = 2000 TeV) 
is needed to prepare for the 
Auger experiment \cite{Auger} as well as for a reliable interpretation 
of present experiments like 
AGASA \cite{Agasa}, Fly's Eye \cite{Flyseye} and High Resolution Fly's
Eye \cite{Abu-Zayyad00b,Abu-Zayyad00a} , 
which present data in the EeV energy region.

Nearly all event generators for hadronic or nuclear collisions 
in the considered energy range are constructed out of  
the following ingredients:

(i)In hadron--hadron collisions or in elementary nucleon--nucleon 
collisions of nuclear collisions multiple soft color singlet chains 
 with quarks, 
diquarks and antiquarks at their ends 
are formed according to the rules 
of Gribov--Regge theory together with topological arguments.
This are mainly diquark--quark and quark--antiquark 
chains. The Gribov--Regge theory also gives all the hadron--hadron 
cross sections.

(ii) In collisions involving nuclei one starts with the construction
of multiple interactions according to the Gribov--Glauber theory.
In simple models each collision leads just to the formation of one pair of 
colorless chains. In evolved models there is the full multi--chain system 
according to (i) and (iii) in each elementary collision.
The Gribov--Glauber theory gives not only the distribution in the 
numbers of elementary collisions, it gives also the nuclear 
cross--sections in terms of the elementary (input) 
nucleon--nucleon cross--sections.
There are also simplified models, 
which do not treat the full Gribov--Glauber theory 
for nucleus--nucleus collisions. In these models (using the superposition 
model) the nucleus--nucleus collision is replaced by a couple of hadron--nucleus
collisions. 

(iii) At sufficiently large energy besides the soft chains according to (i) 
and (ii) we have a system of multiple minijets.
At high collision energies hard and semihard parton--parton collisions
occur and become more and more important with rising collision energy.
The produced jets and minijets become important features of most models,
they determine the transverse momentum or transverse
energy properties of the models at high energy.
The input cross section 
 for semihard multiparticle
production (or minijet production)  $\sigma_{h}$ is calculated applying
the QCD improved parton model,
the details are given (for instance as implemented in DPMJET) 
  in Ref.\cite{CTK87,DTUJET92a,DTUJET92b,Hahn90}.
\begin{eqnarray}
\sigma_h &=& \sum_{i,j,k,l}
\int_0^1 dx_1 \int_0^1
dx_2 \int d\hat{t}\ 
 \frac{1}{1+\delta_{kl}} 
\frac{d\sigma_{QCD,ij \rightarrow kl}}{d\hat{t}}\ 
\nonumber\\
& &\times
f_i(x_1,Q^2)
f_j(x_2,Q^2)\ \Theta(p_\perp-p_{\perp{thr}})
\end{eqnarray}
$f_i(x,Q^2)$ are the structure functions of partons with the flavor
$i$ and scale $Q^2$ and the sum $i,j,k,l$
runs over all possible flavors.
Now after the HERA measurements of structure 
functions at small Bjorken $x_{bj}$
 it is very essential to use modern structure functions, which agree to
 the HERA data.
Since the HERA measurements, the structure functions are known
to behave at small $x$ like 1/$x^{\alpha}$ with $\alpha$ between
1.35 and 1.5.
To remain in the region where perturbation theory is valid, 
 a low $p_{\perp}$ cut--off, $p_{\perp_{thr}}$,  is used 
for the minijet component. The multiplicity distribution for the minijets 
is obtained by an unitarization procedure.

(iv)After forming all partonic chains 
according to (i) to (iii) the next step
is initial state and final state QCD evolution 
of all hard chains, after this
 all chains are hadronized using the Lund codes JETSET or PYTHIA or 
in some codes private 
dedicated codes for string hadronization.

(v)There are special cases of hadron production in hadron--hadron, 
hadron--nucleus and nucleus--nucleus collisions, these are diffractive
interactions (or interactions with rapidity gaps between the produced 
particles). Diffractive hadron production is  
usually treated in special 
routines. Since the discovery of hard processes also in diffractive 
interactions  at the CERN-SPS collider, HERA and at the TEVATRON 
(for instance jet--gap--jet events) this has become more 
complicated than before, only very few codes (for instance PHOJET 
and with this DPMJET--III) include at present hard diffraction. However,
there exist dedicated codes for the simulation of hard diffraction.

(vi) After steps (i) to (v) we have events consisting out 
of produced hadrons
(from string fragmentation) and spectators (these are nucleons from the
original projectile and target nuclei, which have not taken part in the 
interaction). In the next step the codes treat secondary interactions. 
This is in 
simple cases just a formation zone intranuclear cascade, 
where the produced hadrons interact with the spectators, a more complete 
treatment is  the full secondary hadron cascade, where the produced hadrons 
interact with the spectators as well as with other produced co--moving 
hadrons. There are 
also codes, where the secondary interactions occur already at the level
of the chains before hadronization or at the level of chain--end partons before
 the formation of the color neutral chains.

(vii)Now we have the produced hadrons and the remains of the projectile and 
target nucleus (in the form of spectator nucleons with no interactions 
or only interactions which were not able to knock them out of the nuclear 
potential). These remains are considered as 
excited residual nuclei, which in a following 
evaporation step deexcite by the emission of evaporation particles 
(protons, neutrons and nuclear fragments). Finally the excited nuclei, 
which no longer are able to emit evaporation particles emit deexcitation 
photons to form finally a stable (or radioactive) residual nucleus.

Let us summarize  important aspects of some event generators used 
 often in calculating the Cosmic Ray cascade:

 The DPMJET--II.5
event generator based on the two--component Dual Parton Model
(DPM) was described in detail 
 \cite{DPMJETII,Ranftsare95,Dpmjet23,Ranft99a,Ranft99b,Ranft20001}.
  DPMJET--II.5 
 describes well minimum bias hadron and hadron jet
 production up to  present collider energies. 
 DPMJET is used for the simulation of the Cosmic Ray cascade
 within the HEMAS--DPM code \cite{DPMBFR94} used mainly for the
 MACRO experiment \cite{Macro}. DPMJET--II.5 will soon be available
 within the CORSIKA cosmic ray cascade code\cite{Corsika}.

 The SIBYLL--1.7 model \cite{SIBYLL} is a minijet model and has been
 reported to be applicable up to $E_{lab}$ = 10${}^{20}$ eV.
 However, the EHLQ \cite{EHQL} parton structure functions used
 for the calculation of the minijet component might , after the
 HERA experiments, no longer be adequate. It is known, that a
 significant updating of SIBYLL is on the way \cite{Engel99c}, 
 but the new code is not
 yet available.
 SIBYLL is a popular model for simulating the Cosmic Ray
 cascade in the USA. SIBYLL is available within the CORSIKA cosmic ray
 cascade code\cite{Corsika} and within the AIRES \cite{AIRES} and MOCCA
 \cite{MOCCA,MOCCA2} cosmic ray cascade codes .

 VENUS \cite{VENUS}, a  popular model applied originally mainly for describing
 heavy ion experiments. 
 VENUS is applicable up to $E_{lab}$ = 5$\times10{}^{16}$ eV.
 It has been reported \cite{Werner96}, that the introduction of
 minijets into VENUS has been planned, this will allow to apply
 VENUS up to higher energies.  
  A new version of VENUS called {\it neXus}\cite{neXus}
       was not yet to be available for distribution at the time, when my
       talk was prepared, it will be available in the next release of
       the CORSIKA cosmic ray cascade code.

 QGSJET \cite{QGSJET} is the most popular Russian event
 generator used for Cosmic Ray simulations. It is based on the
 Quark Gluon String (QGS) model, this model is largely
 equivalent to the DPM. QGSJET also contains a minijet component
 and is reported to be applicable up to $E_{lab}$ = 10${}^{20}$
 eV. QGSJET is available within the CORSIKA cosmic ray cascade
 code\cite{Corsika} and within the AIRES \cite{AIRES} cosmic ray cascade
 code.

In Table 1 we present some characteristics of the models. The
Gribov--Regge theory is applied by three of the models. The
pomeron intercept for SIBYLL is equal to one, SIBYLL is a
minijet model using a critical pomeron, with only one soft chain
pair, all the
rise of the cross section results from the
minijets. In the models with pomeron intercept bigger than one,
we have also multiple soft chain pairs, already the soft pomeron
leads to some rise of the cross sections with energy. Minijets
are used in three of the models, it is believed, that minijets
are
necessary to reach the highest energies. All models contain
diffractive events. Secondary interactions between all produced
hadrons and spectators exist only in VENUS, DPMJET has only a
formation zone intranuclear cascade (FZIC) between the produced
hadrons and the spectators. Only three of the models sample
properly nucleus--nucleus collisions, SIBYLL
replaces this by the semi--superposition model \cite{EngelJ}. 
The residual projectile (and target)
nuclei are only given by two of the models.

{\bf Table 1.}
Characteristics of some popular models for hadron production  in
Cosmic Ray cascades.  
\vskip 5mm
\begintable
 {}           |VENUS | QGSJET | SIBYLL |  DPMJET   \cr
 Gribov--Glauber | x  | x   |  x  |   x    \cr
 Gribov--Regge | x  | x   |     |   x    \cr
 Pomeron intercept     |1.07|1.07 | 1.00|  1.05  \cr
 minijets     |    |  x  |  x  |   x    \cr
 diffractive events   | x  |  x  |  x  |   x    \cr
 secondary interactions    | x  |     |     |   x    \cr
 A--A interactions    | x  |  x  |     |   x    \cr
 superperposition model      |    |     |  x  |        \cr
 nuclear evaporation, residual nuclei   |    |  x  |     |   x    \cr
 maximum Energy [GeV] |$10^7$ | $10^{11}$ |$10^{11}$ | $10^{12}$  \endtable

Each model has to determine its free parameters.
  This can be done by a global fit to all available data of
total, elastic, inelastic, and single diffractive cross sections in the
energy range from ISR to collider experiments as well as to the
data on the elastic slopes in this energy range.
Since there are some differences
in the hard parton distribution functions at small
$x$
values resulting in different hard input cross sections we have
to perform separate fits for each set of
parton distribution functions.
After this stage each model predicts the cross sections also
outside the energy range, where data are available.

\section{Code comparisons}

There is certainly the need to understand the systematic errors of the
hadronic models and event generators. (i) This is more easy in all the
region, where experimental data exist. Here it is useful to compare the
models to as many experimental data as possible. It is a task of the
community to decide about the most suitable benchmark experiments.
We can however assume, that most published models agree to most of the
available data. 

Systematic errors
of models are more difficult to estimate in regions, where no
experimental data are available. This might be collisions at energies
beyond the energies of existing accelerators or in regions of the phase
space, where no experimental data are available. A very prominent
example for the latter is the fragmentation region at large Feynman--x,
which was never studied in hadron production experiments 
at the highest energies
available   at the CERN--SPS collider and at the TEVATRON
 collider, but just this fragmentation region is of upmost
importance for the Cosmic Ray cascade. 
The systematic comparison
of Monte Carlo models 
 in such regions is very useful to understand how
reliable the extrapolations of the models into such regions might be.

    The Karlsruhe code
    comparison \cite{Kcodcomp} was a first extensive comparison of Monte
    Carlo models, which are available as event generators within the
    CORSIKA Cosmic Ray cascade code  \cite{Corsika,Corsika561}.
    The energies, mainly in the 10$^{14}$ 
    and 10$^{15}$ eV range and the 
    distributions chosen in this comparison were motivated
    by the interest of the KASKADE\cite{Kaskade} 
    experiment in Karlsruhe. 
With one exception, I will not review this code comparison here. A code
comparison extending up to higher energies was performed within the
AIRES \cite{AIRES} code system \cite{AIRESCC}.

Further code comparisons should be strongly encouraged. Some extensions
to higher energy of such a code comparison were presented in talks 
by Engel \cite{Engel99a,Engel99b}.

Here I will present some code
comparisons up to the highest energies, for which the models are able to
run. The first idea was just to use the event generators implemented in
the last distributed version CORSIKA--5.61 \cite{Corsika561}. But there
was only the older version 
DPMJET--II.4 implemented.
 So I use for DPMJET my stand--alone version of DPMJET--II.5
\cite{Ranft99a,Ranft99b,Ranft20001}. With SIBYLL--1.6 as implemented in
CORSIKA--5.61 I got funny results at high energy (above 10$^{18}$ eV) (I
understand, that these problems with SIBYLL--1.6 are now corrected in
CORSIKA.).
Using the stand--alone code SIBYLL--1.7, 
the last SIBYLL version distributed
by the authors, these anomalies disappeared, 
 therefore I
use for SIBYLL only the stand--alone version mentioned. With QGSJET in
CORSIKA, I did not find any problems, so I use it in the form as
implemented in CORSIKA--5.61. The VENUS code as implemented in
CORSIKA--5.61 runs only up to 10$^{16}$ eV. With this it is not possible
to go beyond the Karlsruhe code comparison \cite{Kcodcomp}.

In all the following plots I  present and compare the result of the three
event generators in the implementations as mentioned. The first comparisons are for p--N (proton--Nitrogen) collisions, p--N collisions are a good approximation to p--Air collisions which are important for the Cosmic Ray cascade.

\begin{figure}[thb]
\begin{center}
 \includegraphics[width=5.5cm,height=4.7cm]{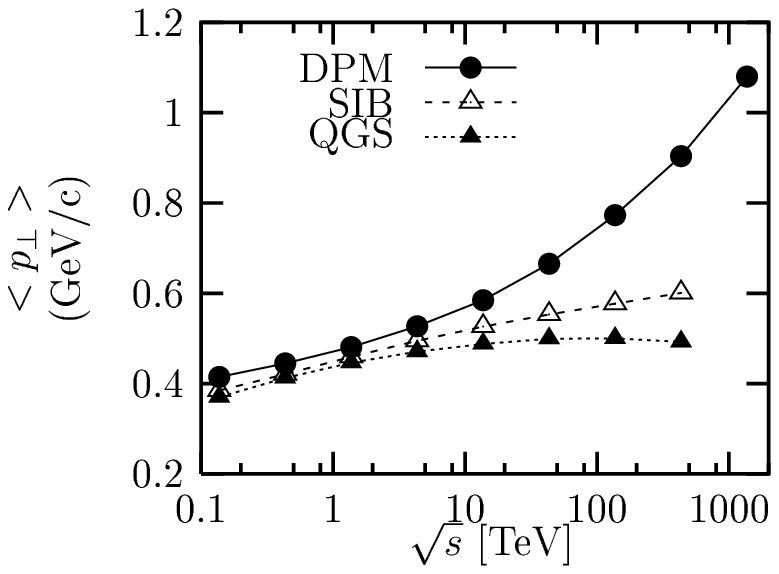}
 \includegraphics[width=5.5cm,height=4.7cm]{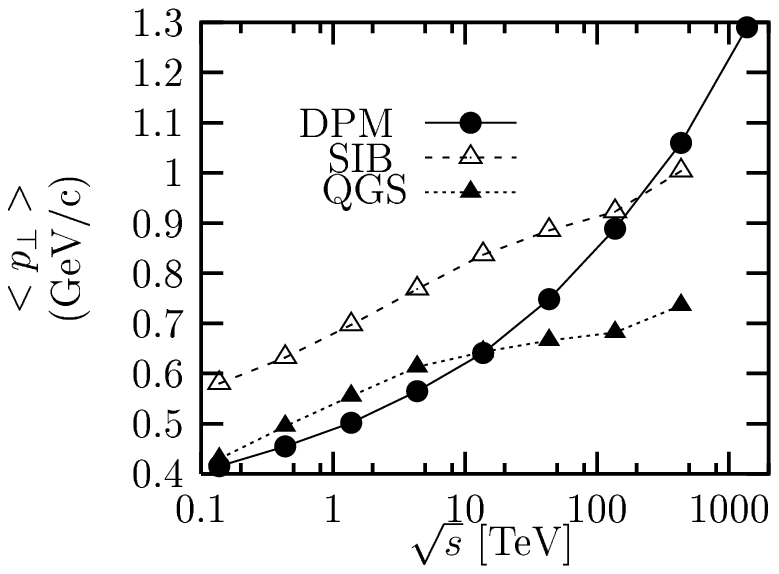}
 \end{center}
 \vspace*{-3mm}
 \caption{{\bf (a)}
 Average transverse momenta  for
 charged hadron  production in p--N  (proton--Nitrogen)
 collisions as function of the (nucleon--nucleon)
cms energy  $\sqrt s$.
 {\bf (b)}
Average transverse momenta  for 
proton  production in p--N (proton--Nitrogen) 
 collisions as function of the (nucleon--nucleon)
cms energy  $\sqrt s$.
\protect\label{cocoptav1}
}
  \end{figure}

In Fig.\ref{cocoptav1}.a we present average 
transverse momenta of charged
hadrons as
obtained from DPMJET--II.5, QGSJET and SIBYLL for p--N collisions
as function of the cms energy $\sqrt s$.  
At energies where data  (in p--p or p--$\bar p$ collisions)
exist
all models agree rather well
with each other and with the data. However, we find completely different
extrapolations to higher energies. We should note, all
 three models have a minijet component. But it seems, that
in spite of the minijets the average transverse momentum in
QGSJET becomes constant at high energies, while it continues to
rise in DPMJET. 
We can conclude, there are very big differences in implementing
the minijet components in the models.
There are even bigger differences in the models, if we consider the
average transverse momenta of secondary protons, see
Fig.\ref{cocoptav1}.b. DPMJET and QGSJET diverge like in
Fig.\ref{cocoptav1}.a and the prediction of SIBYLL--1.7 is even at lower
energies, where we find agreement between all three models in
Fig.\ref{cocoptav1}.a, significantly outside the other models.
\begin{figure}[thb]
\begin{center}
 \includegraphics[width=5.5cm,height=4.7cm]{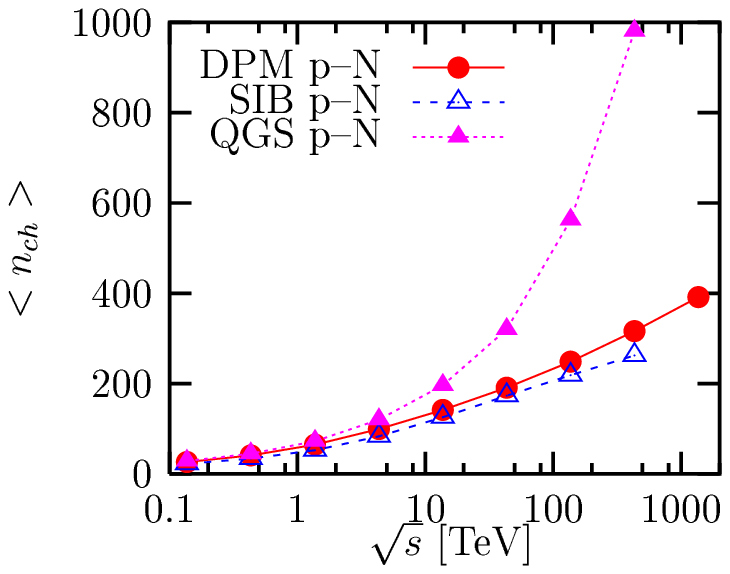}
 \includegraphics[width=5.5cm,height=4.7cm]{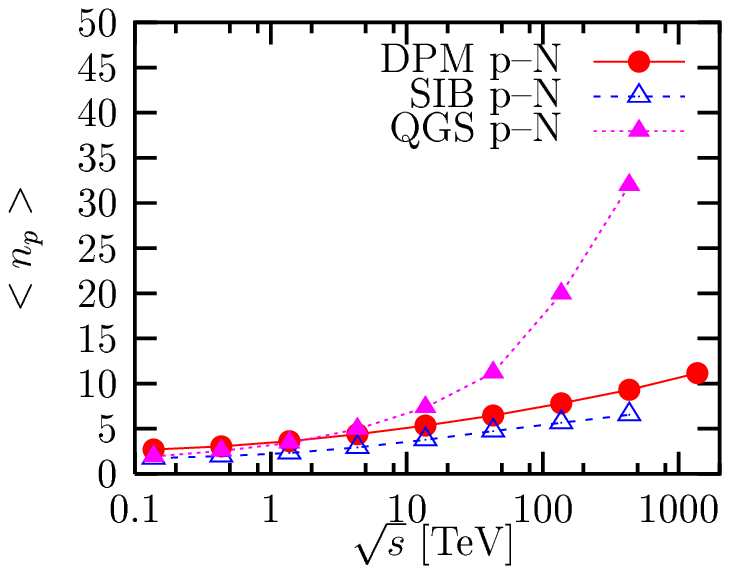}
\end{center}
\vspace*{-3mm}
\caption{{\bf (a)}
The average multiplicity of 
charged hadron  production in p--N (proton--Nitrogen) 
 collisions as function of the (nucleon--nucleon)
cms energy  $\sqrt s$.
{\bf (b)}
The average multiplicity of 
secondary proton  production in p--N (proton--Nitrogen) 
 collisions as function of the (nucleon--nucleon)
cms energy  $\sqrt s$.
\protect\label{coconchpair1}
}
\end{figure}

Fig.\ref{coconchpair1}.a
presents the rise of the total charged  multiplicity and
Fig.\ref{coconchpair1}.b presents the rise of the secondary proton
multiplicity
with the
cms energy $\sqrt s$ according to DPMJET, QGSJET and SIBYLL. We
find again, at low energies, where data are available, the
models agree rather well. DPMJET and SIBYLL agree in all the
energy range shown. However, QGSJET above the energy of the
TEVATRON extrapolates to higher energies in a completely
different way.

\begin{figure}[thb]
\begin{center}
 \includegraphics[width=5.5cm,height=4.7cm]{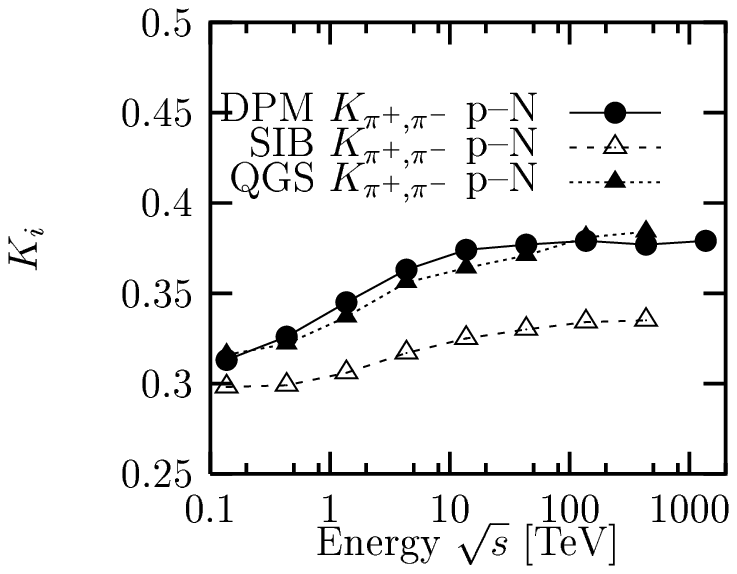}
 \includegraphics[width=5.5cm,height=4.7cm]{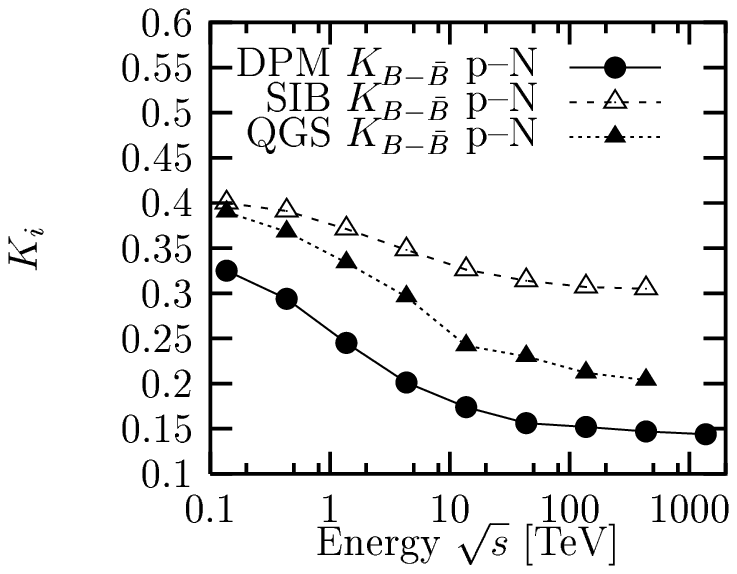}
\end{center}
\vspace*{-3mm}
\caption{{\bf (a)}
Average energy fraction for  charged pion $\pi^+ , \pi^-$ production 
$K_{\pi^+ , \pi^-}$ 
 in p--N 
 collisions as function of the (nucleon--nucleon)
cms energy  $\sqrt s$.
{\bf (b)}
Average energy fraction for  net baryon $B - \bar B$ production 
$K_{B - \bar B}$ 
 in p--N 
 collisions as function of the (nucleon--nucleon)
cms energy  $\sqrt s$.
\protect\label{cocokpipair1}
}
\end{figure}

In Fig.\ref{cocokpipair1}.a and \ref{cocokpipair1}.b  
we present  for  p--N
collisions the energy fractions K  of charged pion production and
for net--baryon $B-\bar B$ production (Since
newly produced baryons and anti--baryons are expected to carry the same
energy fractions, this corresponds to the energy carried by the baryons
which were present before the collision, which have been stopped and
transformed by the collision.).
   In the case of  $K_{B-\bar B}$ we find
  all three models to behave widely different, in the case of 
$K_{\pi^+ , \pi^-}$  we find DPMJET and QGSJET to agree largely, but
SIBYLL  behaves quite differently. Here I should mention, that new
baryon stopping diagrams \cite{Kharzeev96,Capella96} have so far only
been implemented in DPMJET--II.5. This enhanced stopping is responsible
for $K_{B-\bar B}$ being smaller and  $K_{\pi^+ , \pi^-}$ being bigger
in DPMJET--II.5 than in the other two models.

The cosmic ray spectrum--weighted moments \cite{gaistext} 
in p--A collisions 
are defined as moments of the $F(x_{lab}) =x_{lab}dN/dx_{lab}$ :
\begin{equation}
f^{p-A}_i = \int^{1}_0 (x_{lab})^{\gamma -1}
F^{p-A}_i(x_{lab})dx_{lab}
\end{equation}
Here $-\gamma \simeq$ --1.7 is the power of the integral cosmic
ray energy spectrum and $A$ represents both the target nucleus name
and its mass number.

\begin{figure}[thb]
\begin{center}
 \includegraphics[width=5.5cm,height=4.7cm]{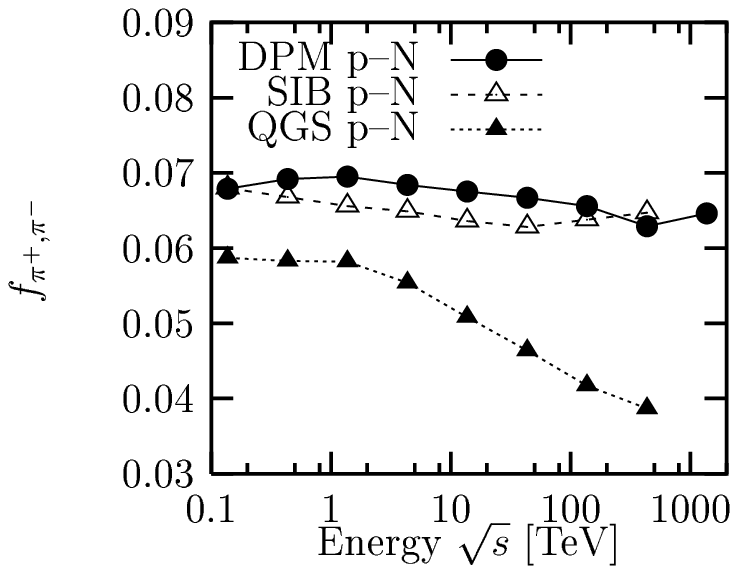}
 \includegraphics[width=5.5cm,height=4.7cm]{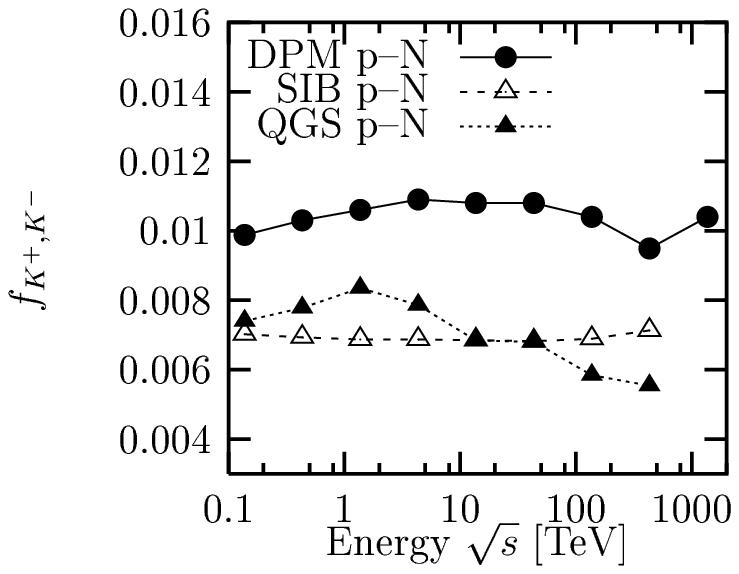}
\end{center}
\vspace*{-3mm}
\caption{{\bf (a)}
 Spectrum weighted moments for charged pion production 
$f_{\pi^+ , \pi^-}$ 
 in p--N 
 collisions as function of the (nucleon--nucleon)
cms energy  $\sqrt s$.
{\bf (b)}
 Spectrum weighted moments for charged Kaon production 
$f_{K^+ , K^-}$ 
 in p--N 
 collisions as function of the (nucleon--nucleon)
cms energy  $\sqrt s$.
\protect\label{cocofpipair1}
}
\end{figure}

 In Fig.\ref{cocofpipair1}.a and \ref{cocofpipair1}.b 
 we present the spectrum weighted moments
 for pion and Kaon production in  p--N collisions as function of
 the cms energy $\sqrt s$ per nucleon. 
In contrast to Fig.
\ref{cocokpipair1} we find this time DPMJET and SIBYLL to agree
approximately, but QGSJET behaves completely different.

\begin{figure}[thb]
\begin{center}
\mbox{
 \includegraphics[width=5.5cm,height=4.7cm]{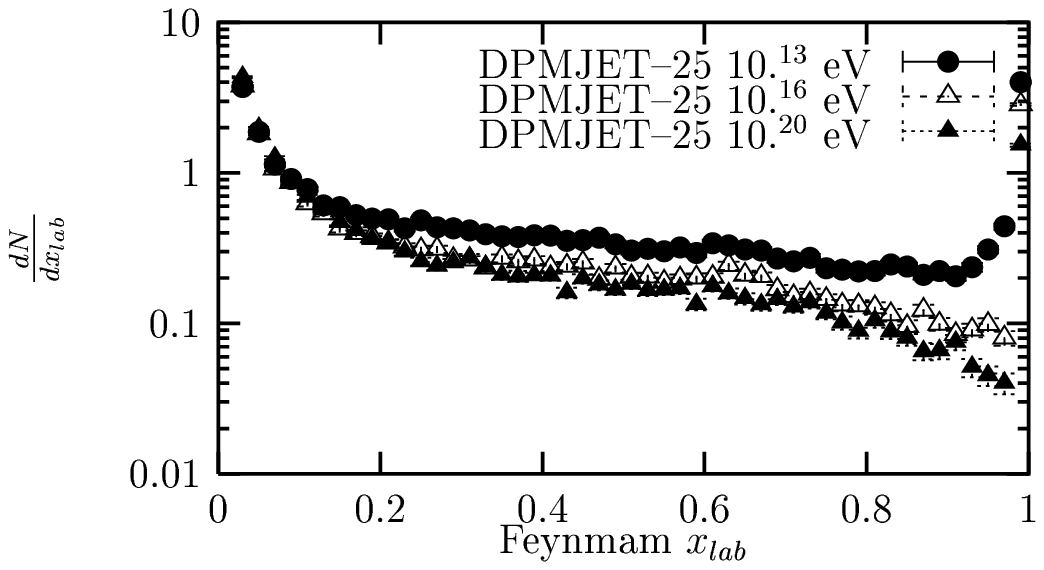}
 \includegraphics[width=5.5cm,height=4.7cm]{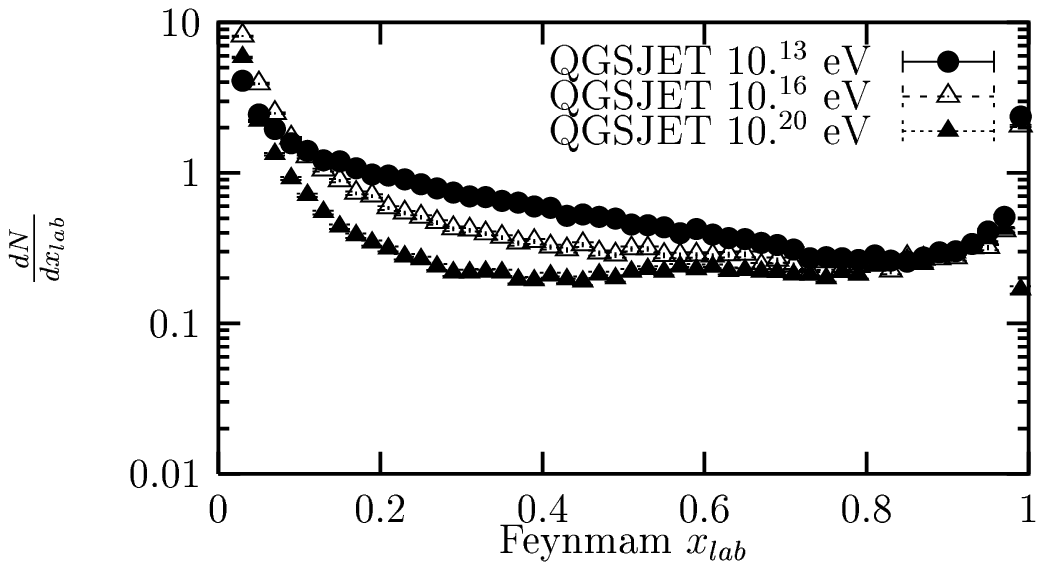}
}
\end{center}
\vspace*{-3mm}
\caption{{\bf (a)}
$x_{lab}$--distributions of secondary protons 
produced in p--N collisions at 
$E_{lab}$ = 10$^{13}$, 10$^{16}$ and 10$^{20}$  eV 
according to the DPMJET--II.5 model.
{\bf (b)}$x_{lab}$--distributions of secondary protons 
produced in p--N collisions at 
$E_{lab}$ = 10$^{13}$, 10$^{16}$ and 10$^{20}$  eV 
according to the QGSJET
 model.
\protect\label{dpm25xlabp1}
}
\end{figure}

In Figs. \ref{dpm25xlabp1}.a and \ref{dpm25xlabp1}.b we present 
the extrapolations to high energy $E_{lab}$ of the  $x_{lab}$ distributions
 $dN/dx_{lab}$ for secondary protons produced according to the 
 models DPMJET--II.5 
 and QGSJET. The distributions are plotted for
 $E_{lab}$ = 10$^{13}$, 10$^{16}$ and 10$^{20}$  eV . At the lowest energy
the two models agree roughly, but there are considerable differences in 
the extrapolations to 10$^{16}$ and 10$^{20}$  eV. The differences between
DPMJET--II.5 and SIBYLL--1.7 are mainly due to the baryon stopping diagrams
implemented in DPMJET--II.5 \cite{Ranft20001,Ranft20003} 
but not present in SIBYLL--1.7. The scaling behavior of the distributions 
according to DPMJET--II.5 and SIBYLL--1.7 is nearly perfect in 
the $x_{lab}$ region from 0.1 to 0.6, but this is just the region where 
the non--scaling appears in QGSJET. The behavior of the models in p--p and 
p--$\bar p$ is rather similar to the one in p-N collisions. Clearly, 
measurements of leading baryon distributions at the TEVATRON and  at the LHC
will resolve the conflict between the three models.

\begin{figure}[thb]
\begin{center}
\mbox{
 \includegraphics[width=5.5cm,height=4.7cm]{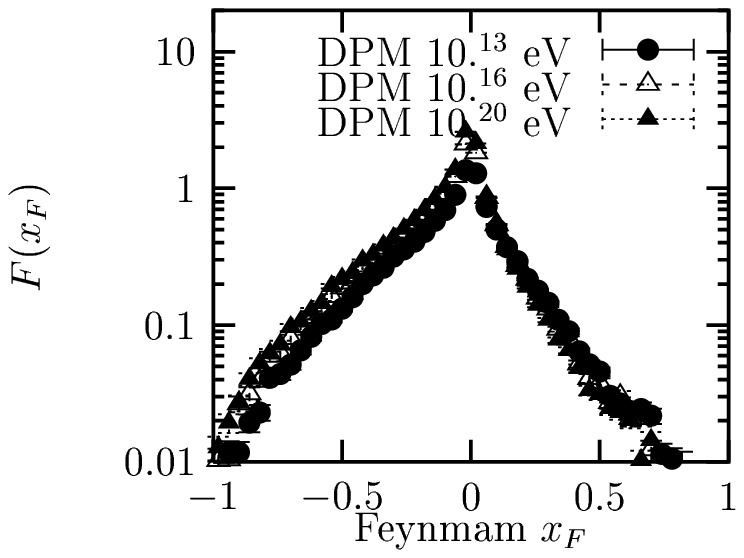}
 \includegraphics[width=5.5cm,height=4.7cm]{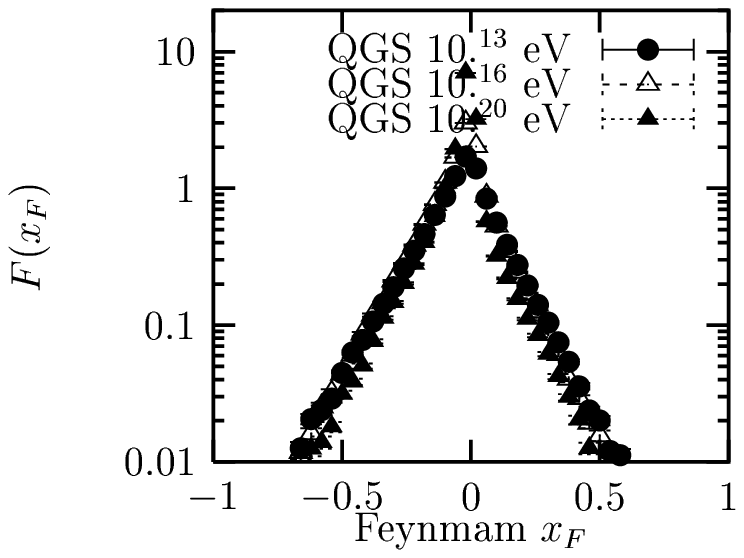}
}
\end{center}
\vspace*{-3mm}
\caption{{\bf (a)}
$x_{F}$--distributions of secondary $\pi^+$ 
produced in p--N collisions at 
$E_{lab}$ = 10$^{13}$, 10$^{16}$ and 10$^{20}$  eV 
according to the DPMJET--II.5 model.
{\bf (b)}$x_{F}$--distributions of secondary $\pi^+$ 
produced in p--N collisions at 
$E_{lab}$ = 10$^{13}$, 10$^{16}$ and 10$^{20}$  eV 
according to the QGSJET
 model.
\protect\label{dpm25xfpip1}
}
\end{figure}

Similarly, in Figs. \ref{dpm25xfpip1}.a and \ref{dpm25xfpip1}.b we present 
the extrapolations to high energy $E_{lab}$ of the  $x_{F}$ distributions
 $F(x_F)$ for secondary $\pi^+$ produced according to the 
 models DPMJET--II.5 
 and QGSJET. The distributions are again plotted for
 $E_{lab}$ = 10$^{13}$, 10$^{16}$ and 10$^{20}$  eV . 
 SIBYLL--1.7 behaves similar to DPMJET--II.5 but as seen before
 DPMJET--II.5 and QGSJET differ considerably in the shape of the
 distributions, only measurements at the
 highest possible energies can resolve the situation.

In the next four Figures we will demonstrate, 
that the differences in the 
extrapolations to high energy in Fe--N collisions 
are at least as large  
as in p--p or p--N collisions. However we find also, that in
Fe--N collisions QGSJET
does not run up to such high energies as DPMJET and SIBYLL.

 \clearpage

For a better comparison with our p--N Figures, we plot  also the Fe--N
comparisons up to $\sqrt s$ = 1000 TeV per nucleon--nucleon collision.
What is needed for cosmic ray simulations are lab energies up to 
10${}^{20}$ to 10${}^{21}$ eV  per nucleus, the energies reached with
QGSJET are sufficient for this. 

\begin{figure}[thb]
\begin{center}
 \includegraphics[width=5.5cm,height=4.7cm]{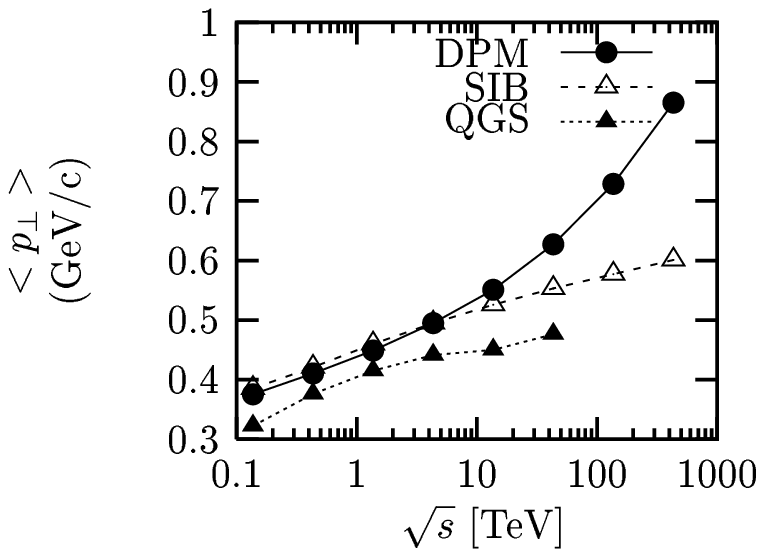}
 \includegraphics[width=5.5cm,height=4.7cm]{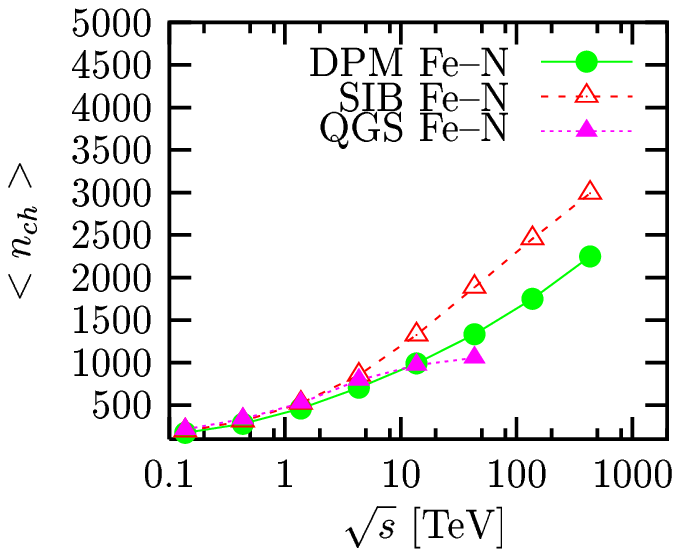}
\end{center}
\vspace*{-3mm}
\caption{{\bf (a)}
Average transverse momenta  for 
charged hadron  production in Fe--N 
 collisions as function of the (nucleon--nucleon)
cms energy  $\sqrt s$.
{\bf (b)}
Average multiplicity of  
charged hadron  production in Fe--N 
 collisions as function of the (nucleon--nucleon)
cms energy  $\sqrt s$.
\protect\label{cocoptavfeair1}
}
\end{figure}

In Fig. \ref{cocoptavfeair1}.a we compare 
average $<p_{\perp}>$ of produced 
charged hadrons as function of the 
(nucleon--nucleon) cms energy  $\sqrt s$ 
in Fe--N collisions. 
The differences between the three models look very similar
to the differences found in p--N collisions in Fig.\ref{cocoptav1}.a.
In Fig. \ref{cocoptavfeair1}.b we compare the multiplicity of produced 
charged hadrons as function of the 
(nucleon--nucleon) cms energy  $\sqrt s$ 
in Fe--N collisions. Here the differences between the three models 
look completely different than 
  in p--N collisions in Fig.\ref{coconchpair1}.a.

\begin{figure}[thb]
\begin{center}
 \includegraphics[width=5.5cm,height=4.7cm]{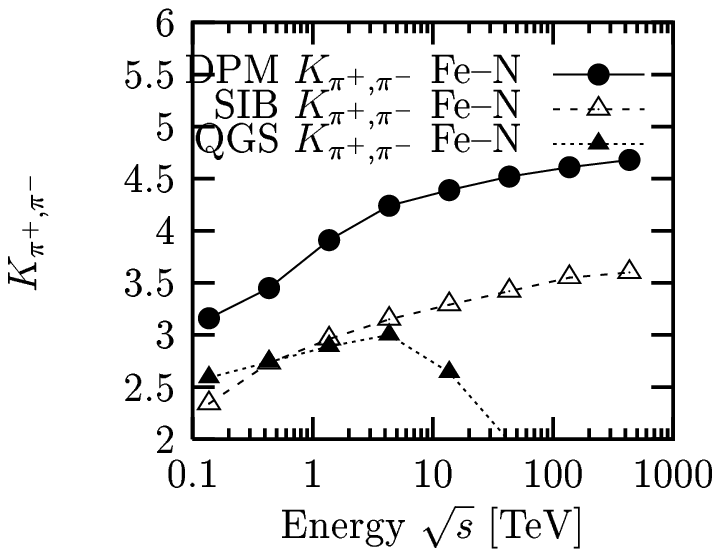}
 \includegraphics[width=5.5cm,height=4.7cm]{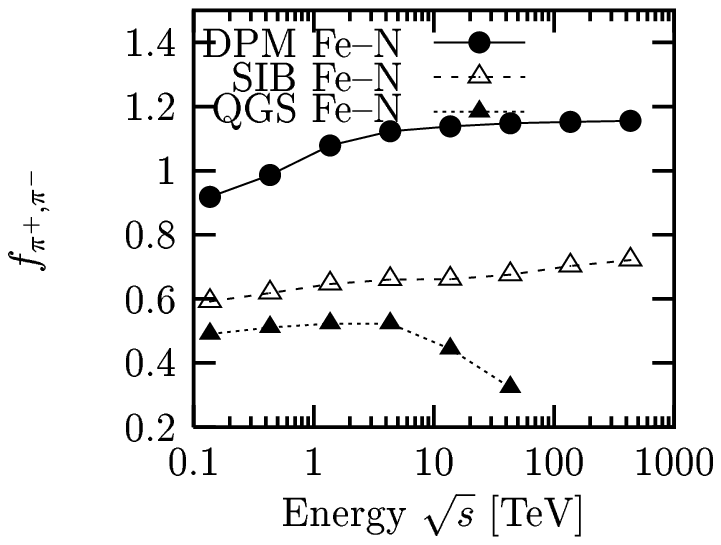}
\end{center}
\vspace*{-3mm}
\caption{{\bf (a)}
Average energy fraction for  charged pion $\pi^+ , \pi^-$ production 
$K_{\pi^+ , \pi^-}$ 
 in Fe--N 
 collisions as function of the (nucleon--nucleon)
cms energy  $\sqrt s$.
{\bf (b)}
 Spectrum weighted moments for charged pion production 
$f_{\pi^+ , \pi^-}$ 
 in Fe--N 
 collisions as function of the (nucleon--nucleon)
cms energy  $\sqrt s$.
\protect\label{cocokpifeair1}
}
\end{figure}

In Fig. \ref{cocokpifeair1},a we compare the 
average energy fraction for  charged pion $\pi^+ , \pi^-$ production 
$K_{\pi^+ , \pi^-}$ 
 in Fe--N 
 collisions as function of the (nucleon--nucleon)
cms energy  $\sqrt s$. At low energies $K_{\pi^+ , \pi^-}$ rises with 
$\sqrt s$ for all three models and 
the absolute magnitudes differ by less than 
20 percent. However, starting at $\sqrt s$ = 5 TeV for QGSJET 
$K_{\pi^+ , \pi^-}$ starts to decrease strongly, 
while for the other two 
models $K_{\pi^+ , \pi^-}$ continues to rise with  $\sqrt s$.

In Fig. \ref{cocokpifeair1}.b we observe a quite similar behavior of  
the spectrum weighted moments for charged pion production 
$f_{\pi^+ , \pi^-}$ 
 in Fe--N 
 collisions as function of the (nucleon--nucleon)
cms energy  $\sqrt s$. Now the curves for DPMJET--II.5 and SIBYLL--1.7 
rise in a similar way with $\sqrt s$ , but the absolute size differs 
considerably. For QGSJET we find , similarly to the last Figure a 
decrease starting at about $\sqrt s$ = 5 TeV.

\section*{Comparison of the models after simulating the Cosmic
 Ray cascade}

\begin{figure}[b]
\begin{center}
\includegraphics[width=8cm,height=7cm]{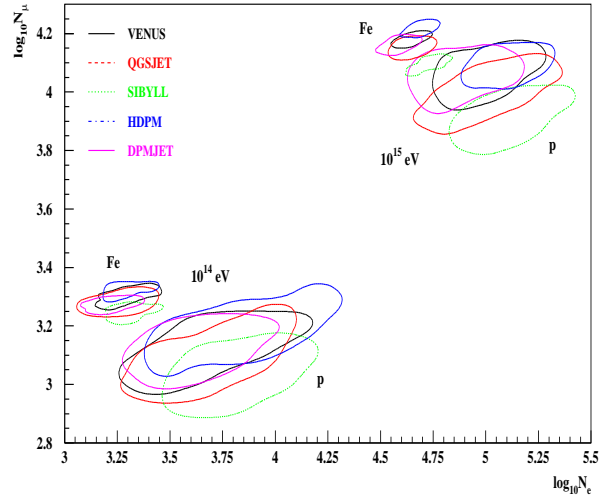}
\end{center}
\caption[]{Contours in the $\log_{10}N_{\mu}$ --$\log_{10}N_{e}$
 plane for p and Fe induced showers of $E = 10^{14}$ and
 $10^{15}$ eV .
 The
 distributions were calculated using the CORSIKA shower code
 \protect\cite{Kcodcomp} with 5 different event generators for the
 hadronic interactions. The HPDM code\cite{HPDM} not mentioned before in this
 talk is a DPM similar model.
 }
\label{knapp3}
\end{figure}

    Next we present one comparisons from the Karlsruhe code
    comparison \cite{Kcodcomp},
 which refers to the Cosmic Ray cascade simulated 
with different event generators.
 The distribution chosen in this comparison is motivated
    by the interest of the KASKADE\cite{Kaskade} 
    experiment in Karlsruhe.

 In Fig. \ref{knapp3} Fe and p induced showers with energies of
 $E = 10^{14}$ and $10^{15}$ eV are plotted in the $\log_{10}
 N_{\mu}$ --$\log_{10}N_{e}$ plane (Muon--number --Electron--number
 plane). The distribution of events according to each of the 5
 interaction models for each energy and primary particle 
 is indicated by contours.
Considering such a plot calculated
with only one of the models, where Muon number is plotted over
electron number, the impression is, that a
simultaneous measurement of Muon--number and Electron number
allows to determine the primary energy as well as the
composition of the primary component. In such a plot we would
see, that for
instance VENUS and DPMJET agree very well, but the contour
according to SIBYLL for Fe projectiles of $E = 10^{15}$ eV
overlaps the VENUS and DPMJET contours for p projectiles at this
energy. From these differences between the models one can
conclude, that at present the systematic errors in the cascade
calculation prevent to identify safely the composition of the primary
component from such measurements.

\section{Standardization of Monte Carlo hadronic production models}

There were in the past and there are at present some activities to
standardize some features of the Monte Carlo codes.
Examples are:

(i)Universal code numbers for particles and nuclei in the
computer codes. The Particle Data Group proposed such a Monte Carlo
particle numbering scheme, see for instance \cite{PDG98}.
This scheme is used for the output of events
by many event generators (examples are PYTHIA, JETSET and other Lund
codes, HERWIG and the DPM event generators DPMJET and PHOJET), 
but the scheme is not suitable for the
internal running
of the codes, therefore each event generator uses a second (internal)
particle numbering scheme. It would certainly be of advantage to
standardize also these internal particle codes.

(ii)A standard COMMON
block for the presentation of the Monte Carlo events was proposed by
working groups for the LEP and LHC colliders.
The structure of this common block has been 
suggested in Refs.~\cite{ZLEP,ZLEP1}.
This standard COMMON
follows closely a scheme first used in Monte Carlo codes of the Lund
group. This COMMON contains not only the final particles produced in the
collisions, it also documents all the internal 
features of the model, for
instance: original projectile and target particles, 
partons, chains formed out of
the partons, hadronization of chains, decay of unstable particles,
intranuclear cascade interactions, excited nuclei, 
evaporation particles and stable residual
nuclei and nuclear fragments.

(iii)A similar effort has been
started by the community of heavy ion
experiments for the event generators for heavy ion collisions \cite{OSCAR}.

\section{Conclusions}

Most hadronic event generators which can be used for simulating hadronic 
and nuclear collisions up to the highest energies are quite similar in
their construction and in the underlying theoretical concepts.
At energies, where data from accelerator and collider experiments are available
the models agree rather well with each other and with the most important
 features of the data. As soon as we compare the extrapolations of the models at
higher energy we find in spite of the similarities in the underlying 
theoretical concepts quite often  striking differences between the 
predictions of the models. We conclude: (i) Measurements of inclusive hadron 
production at the TEVATRON Collider and in the future at the LHC collider
are very important to guide the models. (ii) More theoretical efforts to improve
 the hadronic and nuclear collision models are needed to get better 
extrapolations.  (iii)Cosmic ray experiments should use in the high
energy region  simulations with
more than one hadronic interaction model for interpreting their data.

\section{Acknowledgments}
The author thanks  D.Heck for providing him with the CORSIKA code,
 J.Knapp for Fig. 9 and  R.Engel for providing him with the
SIBYLL--1.7 code and for useful discussions.

%

 \clearpage

%
 \end{document}